\documentclass[aip,jcp,reprint,longbibliography]{revtex4-1}
\usepackage{epsfig}
\usepackage{color}
\usepackage{amsmath}
\usepackage{amsfonts}
\usepackage{natbib}
\usepackage{graphicx}
\usepackage{subfigure}
\usepackage{multirow} 
\usepackage{lipsum}
\usepackage{afterpage}
\usepackage{placeins}
\usepackage{float}
\usepackage{braket}
\usepackage[normalem]{ulem}

\newcommand{\avg}[1]{\left<#1\right>}
\newcommand{\len}[1]{\left|#1\right|}

\newcommand{\para}[1]{\left(#1\right)}

\newcommand{\kT}{\ensuremath{k_{\rm B}T}}
\newcommand{\Mb}{\ensuremath{\mathcal{M}}}

\newcommand{\Zb}{\ensuremath{\mathcal{Z}}}
\newcommand{\Rbar}{{\textbf{R}}}

\newcommand{\vb}{\ensuremath{\mathbf{v}}}

\begin{document}


\title{Dynamic Local Symmetry Fluctuations of Electron Density in Halide Perovskites}


\author{Colin M. Hylton-Farrington}
\author{Richard C. Remsing}
\email[]{rick.remsing@rutgers.edu}
\affiliation{Department of Chemistry and Chemical Biology, Rutgers University, Piscataway, NJ 08854}




\begin{abstract}
Metal halide perovskites have emerged as an exciting class of materials for applications in solar energy harvesting, optical devices, catalysis, and other photophysical applications. 
Many of the exciting properties of halide perovskites are tied to their soft, dynamic, and anharmonic lattice. 
In particular, the precise coupling between anharmonic lattice dynamics and electronic fluctuations is not completely understood.
To build an understanding of this coupling, we use ab initio molecular dynamics simulations supplemented by the calculation of maximally localized Wannier functions to carry out a dynamic group theory analysis of local electron density fluctuations and how these fluctuations are coupled to lattice fluctuations in the model inorganic halide perovskite CsSnBr$_3$.
We detail symmetry-dependent couplings between vibrational modes, including octahedral tilting. 
Importantly, we suggest that the large anharmonicity of some of the vibrational modes in CsSnBr$_3$ result from electron rotation--nuclear translation coupling, in analogy to rotation--translation coupling effects in molecular plastic crystals.
We also identify electronic fluctuations in the Cs cation that couple to distortions in the surrounding Sn-Br cubic coordination environment. 
We anticipate that our approach and resulting insights into electronic fluctuations will aid in further understanding the role of the fluctuating lattice in determining important physical properties of halide perovskites and beyond.
\end{abstract}


\maketitle

\raggedbottom

\section{Introduction}

Traditional approaches for understanding solid-state materials rely on their averaged crystal structures.
However, lost in average structures are local structural distortions of the coupled nuclear and electronic densities. 
Local fluctuations in nuclear or electronic structure can lead to correlations and order/disorder that is not readily resolved in average crystal structures~\cite{aeppli2020hidden,bhowal2022,bhowal2024,Maughan2023,carnevali2023,liu2024,wright2016,huang2022,chandra2014ising,chandra2013hastatic}. 
This hidden disorder can play an important role in determining materials properties.
Of particular interest here is the type of hidden disorder known as emphanisis, in which the local symmetry is broken and an ion displaces away from its average position. 
Ultimately, emphanisis usually results from the localization of electron pairs on an ion in a manner that create a local dipole moment or other multipole that interacts strongly enough with its surrounding to cause it to displace significantly far away from its position in the average crystal structure. 
This was first discovered in lead chalcogenides, where stereochemical expression of the lead $s^2$ lone pair leads to dynamic off-centering of lead ions~\cite{bovzin2010entropically,jensen2012lattice,sangiorgio2018correlated}.
Since this discovery, a wealth of materials exhibiting emphanisis have been identified, suggesting that this phenomenon is more common than originally anticipated, including technologically-relevant metal halide perovskites~\cite{fabini2020,fabini2016dynamic,laurita2022,laurita2017chemical,D3TA05315F}. 
Understanding the hidden disorder in $ABX_3$ halide perovskites is of paramount importance to inform their use in solar energy harvesting, optoelectronics, and catalysis~\cite{miyata2017lead,manser2016intriguing,jena2019halide,mozur2021cation,Zhu:2019aa,D0NR07716J,D3TA05315F}.
Here, $A$ is a monovalent cation, $B$ is a divalent cations, and $X$ is a halide. 
Similar to emphanisis in PbTe~\cite{bovzin2010entropically}, the dynamic off-centering of $B$-site metal ions in $ABX_3$ halide perovskites results from localization of their $s^2$ lone pair to one side due to $s-p$ mixing, which results in a dipole moment on the metal ion that interacts with the surrounding halide coordination shell. 
Dynamic disorder is present in the major phases of halide perovskites~\cite{zhao2020polymorphous,Maity:2022aa}, but is most prevalent in the cubic phase that is stable at high temperatures relative to the orthorhombic and tetragonal phases.
Emphanisis has been linked to many important properties of halide perovskites, such as low thermal conductivities~\cite{Rakita_Cohen_Kedem_Hodes_Cahen_2015,laurita2017chemical,Dutta:2021aa,Ghosh:2022aa}, changes in the average band gap~\cite{fabini2020,fabini2016dynamic,laurita2022,laurita2017chemical}, band gap fluctuations~\cite{remsing2020lone,remsing2020new} that determine decoherence times~\cite{shi2020edge}, and stabilization of bright excitions~\cite{becker2018bright,swift2023}.
Dynamic off-centering is also suggested to impact vibrational dynamics of halide perovskites, including dynamic tilting of the halide octahedra surrounding the $B$-site cation~\cite{gao2021metal}.
However, recent work has challenged the hypothesis that the $B$-site lone pair induces octahedral tilting and demonstrated that perovskites with electronically symmetric $B$-site cations also exhibit octahedral tilting, albeit to a lesser extent than their localized lone pair-containing analogues~\cite{caicedo2023disentangling}.
Nonetheless, there appears to be some form of dynamic coupling between the vibrational modes of halide perovskites and fluctuations of their localized electron density --- those of the $B$-site lone pairs at the very least.
Here, we present a group theory-based approach for quantifying static and dynamic orientational electronic disorder in halide perovskites and focus on cubic CsSnBr$_3$ as a model inorganic halide perovskite of interest.
We construct orientational order parameters from rotor functions that involve symmetry-adapted functions appropriate to the irreducible representations of the site symmetry of each ion type. 
We then use these order parameters to quantify the static and dynamic correlations that exist among nuclear displacements and orientations of the local electron density for both single-site and collective fluctuations.
We uncover strong couplings between electronic orientations of Sn and Br. 
Moreover, we demonstrate that the local electronic structure is coupled to phonon modes like octahedral tilting and suggest that orientational disorder of local electron density is indicative of the presence of soft anharmonic phonon modes.
We also quantify electronic fluctuations of the Cs cation and its cubic coordination environment and find non-trivial couplings between the two. 
We anticipate that these results will be useful in understanding the coupling between local electron density and lattice dynamics that dictate many critical processes in halide perovskites, including polarization fluctuations in charge transport~\cite{schilcher2023correlated,guo2019dynamic,mayers2018lattice,schilcher2021significance} and the overall stability of these materials~\cite{TU20212765}.

\section{Simulation Details}
We performed density functional theory (DFT)-based Born-Oppenheimer ab initio molecular dynamics (AIMD) simulations
of CsSnBr$_3$.
The simulations used the QUICKSTEP electronic structure module within the CP2K ab initio code~\cite{CP2K}.
QUICKSTEP utilizes a dual atom-centered Gaussian and Plane wave (GPW) basis approach for the representation of wavefunctions and electron density, resulting in an efficient and precise implementation of DFT.
The molecularly optimized (MOLOPT) Goedecker-Teter-Hutter (GTH) double-$\zeta$ single polarization short-ranged (DZVP-MOLOPT-SR-GTH) Gaussian basis was selected for the expansion of orbital functions. 
Our simulations employed a plane-wave basis with a cutoff of 350~Ry and a REL\_CUTOFF of 50~Ry to represent the electron density~\cite{vandevondele2007}.
Core electrons were represented using Goedecker-Teter-Hutter (GTH) pseudopotentials~\cite{goedecker1996}.
Exchange correlation interactions were approximated using the Perdew-Burke-Ernzerhof (PBE) generalized gradient approximation for the exchange-correlation functional~\cite{perdew1996}, as implemented in CP2K.
Grimme's D3 van der Waals correction was applied to account for the long-range dispersion interactions in the simulations~\cite{grimme2010}.

Our simulations use a $4\times4\times4$ supercell of CsSnBr$_3$ based on the experimentally determined unit cell~\cite{lunt2018}, resulting in a cubic simulation cell with length $L=23.18$~\AA.
We equilibrated the system for at least 5~ps before performing production runs.
These equilibration simulations propagated the equations of motion with the canonical velocity rescaling (CSVR) thermostat to maintain a constant temperature of 50~K or 300~K in the canonical (NVT) ensemble.
Unless otherwise stated, results are for a temperature of $T=300$~K, where the cubic phase is thermodynamically stable.
For the calculation of dynamic properties, we performed AIMD simulations in the microcanonical (NVE) ensemble with a timestep of 0.5~fs for a total trajectory length of approximately 260~ps at 300~K and 36~ps for 50~K.
Maximally localized Wannier functions (MLWFs)~\cite{marzari2012maximally} and their centers (MLWFCs) were computed on the fly using CP2K. 
The spreads of the MLWFs were minimized according to established methods~\cite{berghold2000}.

\section{Theory}
\subsection{Electron Density}

 \begin{figure}
  {\includegraphics[width=0.48\textwidth]{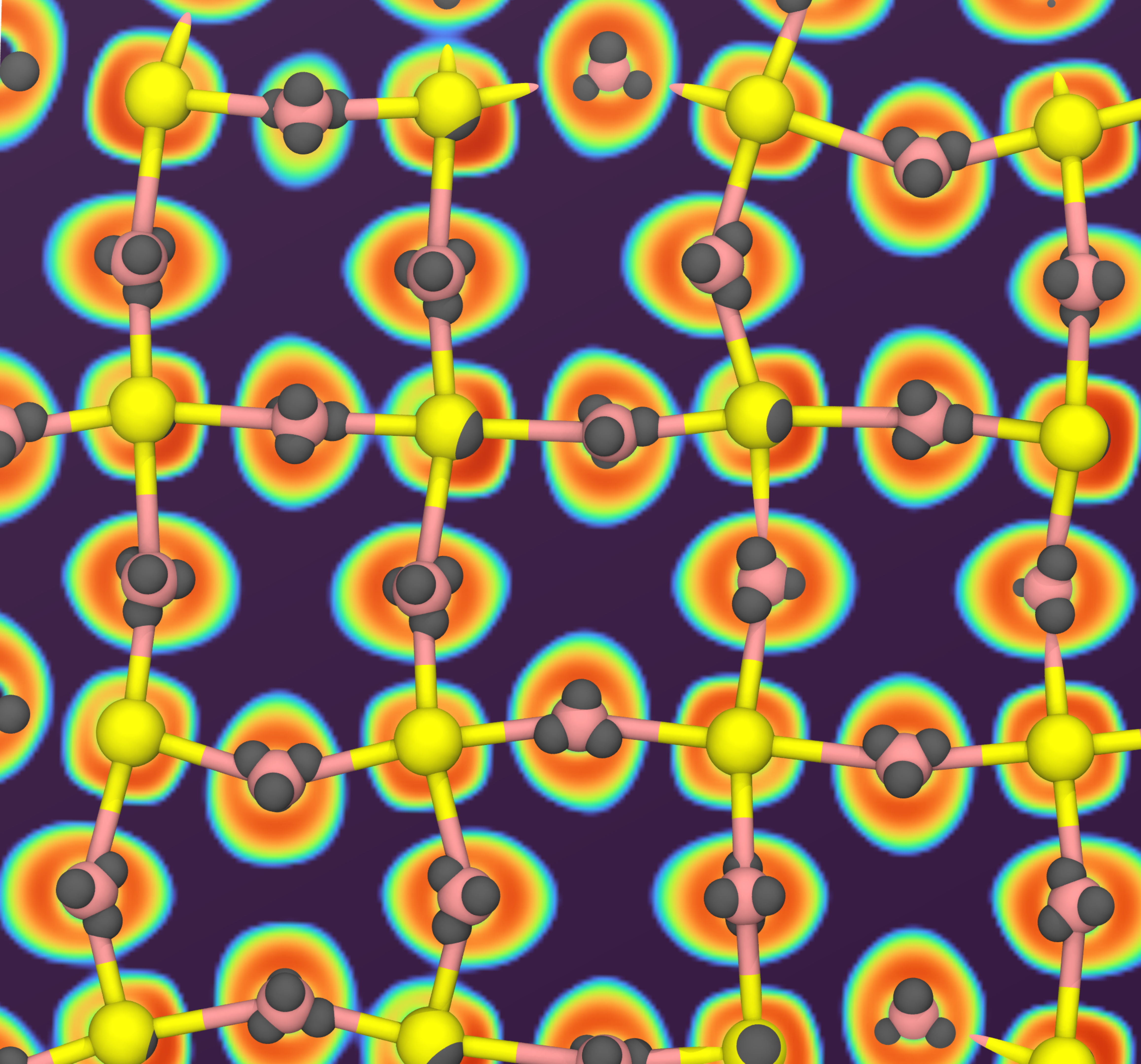}}
  \caption{Contour plot of a slice of the electron localization function (ELF) overlayed onto a snapshot of the configuration of CsSnBr$_3$ for which the ELF was calculated. 
  Also shown are the MLWFCs for each ion in the plane of the ELF slice. 
  Sn are colored yellow, Br are colored pink, and MLWFCs are colored gray.}
  \label{fig:elf}
\end{figure}

%
In this work, we focus on slow electronic modes that are coupled to fluctuations in nuclear positions.
As a result, we work within the Born-Oppenheimer approximation and neglect the fast electronic motion that necessitates a more detailed description of quantum dynamics. 
In periodic systems like those studied here, the eigenstates of the Hamiltonian are the delocalized, periodic Bloch orbitals.
However, local fluctuations in electronic structure are difficult to quantify using Bloch orbitals because they are not localized in real space. 
Because of this, we instead focus our analysis on Wannier functions, which are localized in real space and can be obtained from a unitary transformation of Bloch orbitals.
Wannier functions are not eigenstates of the Hamiltonian, but the charge density of the system is preserved
under the unitary transformation, such that we can use appropriately-defined Wannier functions to quantify
local fluctuations of the electronic density. 
In particular, we focus on MLWFs, which satisfy the non-uniqueness of Wannier functions originating in their gauge freedom by minimizing their spread~\cite{marzari2012maximally}. 
To efficiently quantify fluctuations in electronic structure over the course of AIMD trajectories, we reduce the three-dimensional MLWFs to only their centers.
These MLWFCs provide a reasonably accurate description of electronic charge densities for localized charge distributions, like those studied here. 
Therefore, we use MLWFCs and the changes in their positions along simulation trajectories as a probe of slow local fluctuations in the electronic density that are coupled to the vibrational dynamics of the solid. 
The utility of using MLWFCs to describe local fluctuations in the electron density can be observed by comparing
them to the electron localization function (ELF)~\cite{savin1997elf}.
The ELF is a three-dimensional function (3D), and a two-dimensional (2D) slice is shown in Fig.~\ref{fig:elf} for a representative configuration of CsSnBr$_3$ from our AIMD trajectory.
From this slice, one can identify important features like the localization of Sn lone pairs and their orientation toward the face of the surrounding Br octahedra, evidenced by higher values (darker red) of the ELF. 
Similarly, aspherical density around the Br anions can also be observed. 
However, a complete, time-dependent analysis of this 3D function is complicated by ambiguities associated with defining appropriate isosurface contours, as well as the large storage requirements for these 3D distributions. 
Instead, we focus on MLWFCs, which produce a picture of the electronic structure in agreement with the ELF and similar densities, Fig.~\ref{fig:elf}.
The Sn lone pairs that are pointed toward the face of the octahedron and in the 2D plane of the ELF can be easily seen, as indicated by a gray sphere on a Sn ion oriented in the direction of the high intensity lobe of the ELF in that region. 
Similarly, the MLWFCs for the Br electron pairs closely align with the asphericity of the ELF near the Br. 
Therefore, we expect that fluctuations of the local electron density can be probed by monitoring the dynamic structure of MLWFCs as a proxy for the more complicated full electron density surface. 

\subsection{Rotor Functions}

 \begin{figure*}[tb]
   {\includegraphics[width=0.85\textwidth]{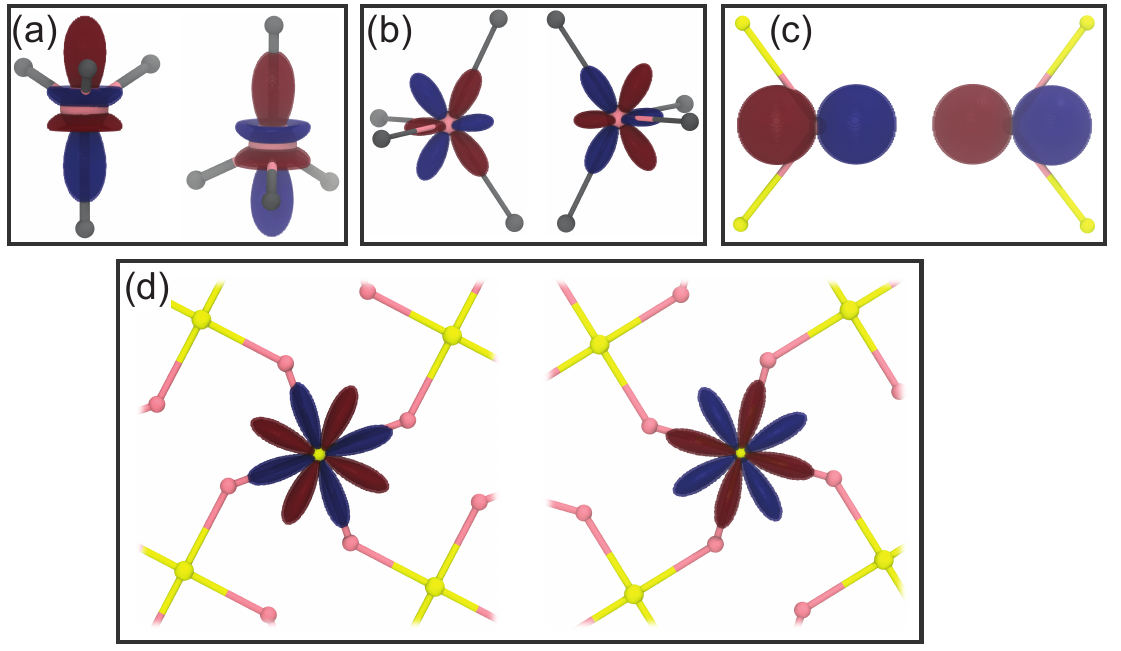}}   
  \caption{Examples of symmetry adapted functions with representative orientations of bond vectors overlapping the function. For each panel, two orientations of opposite sign are shown. We show
  {(a)} the Br--MLWFC ${\mathcal S}_{\mathcal Z}^{3,3}$ ($A_{2u}^{(1)}$) that distinguishes polarization along the local $z$-axis,
  {(b)} the Br--MLWFC ${\mathcal S}_{\mathcal Z}^{3,2}$ ($E_{u}^{(2)}$) that characterizes orientation in the local $xy$-plane, 
  {(c)} the Br-Sn ${\mathcal S}_{\mathcal Z}^{1,0}$ ($E_{u}^{(1)}$) that characterizes to Sn-Br-Sn bending in the $xy$-plane, and
  {(d)} the Sn--Br ${\mathcal S}_{\mathcal K}^{4,3}$ ($T_{1g}^{(1)}$) that characterizes octahedral tilting. 
  The rotor functions are indicated by blue and red isosurfaces for positive and negative values. Sn is colored yellow, Br is pink, and Cs is cyan.}
  \label{fig:rotor_examples}
\end{figure*}

\begin{figure}[ht]
 {\includegraphics[width=0.48\textwidth]{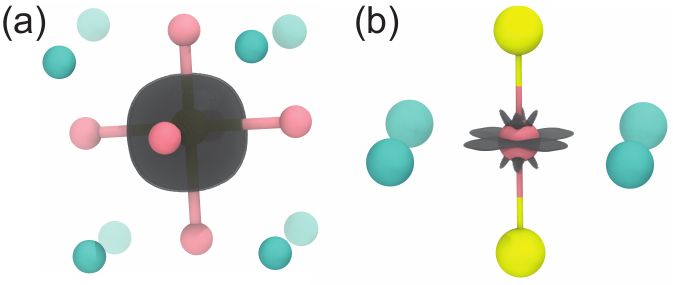}}
	\caption{Average orientational distribution functions, $f_\mu(\Theta)$, of MLWFCs computed from the Laplace series in Eq.~\ref{eq:laplaceseries} from $0 \leq \ell \leq 18$ for
	(a) the Sn--MLWFC vector and (b) the orientation of the four MLWFCs around a Br.
	The average orientational distribution functions are shown as gray isosurfaces. Sn is colored yellow, Br is pink, and Cs is cyan.}
	\label{fig:odf}
\end{figure}

%
To quantify the symmetry of the electron density and atomic structures, we focus on the orientational distribution function of the MLWFCs around a particular ion of interest, $f_{\mu}\para{\Theta}$.
We use rotor functions as order parameters~\cite{lynden1994translation}, which naturally arise in the expansion of the orientational distribution function (derived in the SI),
\begin{equation}
f_{\mu}\para{\Theta}=\frac{1}{4\pi} + 
\sum\limits_{\ell=1}^{\infty}\ \sum\limits_{m=0}^{2\ell} \avg{\mathcal{M}_{\xi}^{\ell,m}(\Rbar;\mu)}
\mathcal{S}_{\xi}^{\ell,m}\para{\Theta},
\label{eq:laplaceseries}
\end{equation}
where $\avg{\cdots}$ indicates an ensemble average over configurations $\Rbar$ and the sums are over angular momentum ($\ell$) and index $m$. 
Note that $0 \leq m \leq 2\ell$. 
The composite label $\mu=(\alpha,\beta)$ defines the vectors for which the solid angle, $\Theta$, is determined. 
The first label, $\alpha$, indicates the identity of the central atom. 
The second label, $\beta$, defines a vector or set of vectors made by $\alpha$ and symmetry equivalent sites in its first coordination shell.
These symmetry equivalent sites can be atoms (in the undistorted crystal structure) or nearest MLWFCs.
For example, $\alpha=$ Br and $\beta=$ MLWFC defines a set of vectors with Br as the central atom and the immediately surrounding four MLWFCs, or $\alpha=$ Sn and $\beta=$ Br indicates the Sn-Br octahedron.
This expansion is similar to the multipole expansion, and the rotor functions describe the structure of the multipoles of increasing angular momentum ($\ell$).
A rotor function, $\mathcal{M}_{\xi}^{\ell,m}$, describes the orientation of vectors centered at a defined site under a symmetry-adapted function $\mathcal{S}_{\xi}^{\ell,m}$ according to
\begin{equation}
{\mathcal M}_{\xi}^{\ell,m} = \frac{1}{N}\sum_{i=1}^{N}{\mathcal S}_{\xi}^{\ell,m}\para{\Theta_{i}},
\label{eq:rotor}
\end{equation}
where {\itshape i} is the index of a vector centered on the site of interest, $N$ is the number of vectors for that site,
and $\Theta_i$ is the solid angle of vector $i$.
Rotor functions have been extensively used to quantify the orientational fluctuations of molecular units in plastic crystals~\cite{michel1985,huller1972}. 
Similarly, rotor functions serve as natural order parameters to quantify the orientational fluctuations of localized electron pairs in ionic solids. 
In crystals, the symmetry-adapted functions $\mathcal{S}_{\xi}^{\ell,m}$ are linear combinations of spherical harmonics that are partner functions to irreducible representations (irreps) of a particular crystallographic point group, where $\xi$ denotes the type of harmonic.
For all but cubic crystallographic point groups, the partner functions are the real spherical harmonics, also known as the Tesseral harmonics, $\xi = \mathcal Z$.
Examples of $\mathcal{S}_{\mathcal{Z}}^{\ell,m}$ are shown in Fig.~\ref{fig:rotor_examples}a,b for units composed of Br and its four MLWFCs, and in Fig.~\ref{fig:rotor_examples}c for Sn-Br-Sn bending. 
Each panel shows two configurations that lead to opposite signs of the symmetry-adapted function.
For cubic point groups, the symmetry-adapted functions are the cubic harmonics, $\xi = \mathcal K$. 
An example of this type of symmetry-adapted function is shown in Fig.~\ref{fig:rotor_examples}d
for Sn-Br correlations that quantify the orientation of four in-plane Br ions around a central Sn. 
The two configurations shown lead to opposite signs of $\mathcal{S}_{\mathcal{K}}^{\ell,m}$,
and, in this case, physically correspond to octahedral tilting in different directions.
Using the rotor functions that result from summing the appropriate symmetry-adapted functions,
we can compute the orientational distribution function according to Eq.~\ref{eq:laplaceseries},
examples of which are shown in Fig.~\ref{fig:odf}
for the MLWFC of tin and the four MLWFCs of bromide.
The symmetry of each distribution reflects the \emph{average} symmetry of the site. 
The tin-MLWFC $f_\mu(\Theta)$ has $O_h$ average symmetry and the bromide-MLWFC $f_\mu(\Theta)$ has $D_{4h}$ average symmetry. 
However, the instantaneous symmetry of each ion-MLWFC unit is incommensurate with the average site symmetry. 
As a result, the orientation of the MLWFCs fluctuates, and the average distribution with the expected site symmetry results from appropriately averaging over the distribution of these electronic fluctuations. 
These orientational electronic fluctuations are the focus of our work.
%

%
Rotor functions with a non-zero value of $\avg{\mathcal{M}_{\xi}^{\ell,m}}$ are partner functions to the fully symmetric irrep of the orientational probability distribution of a system (Eq.~\ref{eq:laplaceseries}).
Rotor functions that belong to other irreps can have non-zero fluctuations due to local dynamic symmetry breaking,
$\avg{(\mathcal{M}_{\xi}^{\ell,m})^2}\neq0$.
Averages of selected rotor functions and their variances are tabulated in Tables~S1, S2, and~S3 for Br-MLWFC, Sn-MLWFC, and Cs-MLWFC units, respectively.
We find that rotor functions with $\ell=3$ are most useful to quantify fluctuations in Br-MLWFC and Cs-MLWFC orientations, while $\ell=1$ is sufficient to quantify Sn-MLWFC orientations, because they are the lowest $\ell$ that results in relatively large fluctuations (see SI). 

\section{Results and Discussion}

\subsection{Symmetry-dependent vibrations in CsSnBr$_3$}

\begin{figure}
  \includegraphics[width=0.48\textwidth]{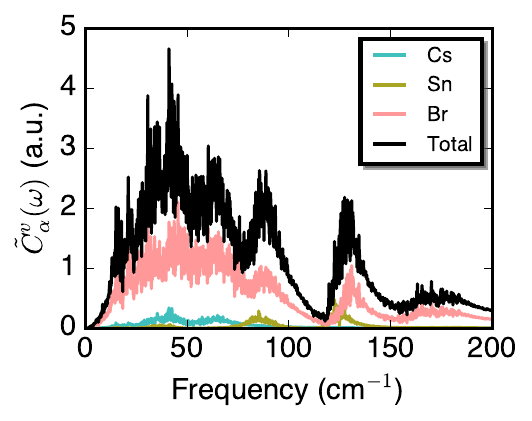}
  \caption{Total vibrational density of states and its atomic contributions from Cs, Sn, and Br.}
  \label{fig:total_vdos}
\end{figure}

Vibrations in CsSnBr$_3$ can be quantified with the phonon density of states, 
\begin{equation}
\tilde{C}^v(\omega) \propto \len{\int_{-\infty}^\infty {\rm d}t C^v(t) e^{-i\omega t}}^2,
\end{equation}
which is computed as the power spectrum of the concentration-weighted velocity autocorrelation function,
\begin{equation}
C^v(t) = \frac{\sum_\alpha c_\alpha \avg{\vb(t;\alpha)\cdot \vb(0;\alpha)}}{\sum_\alpha c_\alpha \avg{\vb^2(0;\alpha)}} = \frac{\sum_\alpha C_{\alpha}^v(t)}{\sum_\alpha C_{\alpha}^v(0)},
\end{equation}
where $c_\alpha$ is the concentration of species $\alpha$, and $\vb(t;\alpha)$ is the velocity of an atom of type $\alpha$ at time $t$.
The total vibrational density of states covers a broad range of frequencies below 200~cm$^{-1}$
and exhibits several peaks over the range of 40 to 170~cm$^{-1}$, Fig.~\ref{fig:total_vdos}.
By splitting $\tilde{C}^v(\omega)$ into its atomic contributions, we see that Cs vibrations occur at frequencies below 100~cm$^{-1}$, as may be expected from its large mass.
Sn vibrations contribute to the peaks at approximately 40~cm$^{-1}$, 90~cm$^{-1}$, and 130~cm$^{-1}$, and Br is involved in all peaks, including being the dominant contributor to the broad peak near 170~cm$^{-1}$.
The atom-specific vibrational densities of states suggest that low frequency modes are dominated by Cs and Br displacements, while high frequency ($>70$~cm$^{-1}$) peaks arise from Br and Sn vibrational modes.
%

%
The overlap in the frequency ranges of Sn and Br atomic vibrational densities of states suggests a coupling between Sn and Br displacements. 
However, Br vibrations can have one of two symmetries, and Sn does not need to couple to both. 
To uncover which vibrational modes of Br and Sn may be coupled, we further decompose the atomic vibrational densities of states based on symmetry.
To do so, we reduce the velocity vector into its irreps, yielding the irrep-dependent velocity autocorrelation function
\begin{equation}
C^{v,\Gamma}_\alpha (t) \propto  \sum\limits_{i \in (x,y,z)}\delta_{\Gamma,\tilde{\Gamma}(i)}  \avg{v_i(0;\alpha) v_i(t;\alpha)},
\label{eq:v_cor}
\end{equation}
where $\tilde{\Gamma}(i)$ returns the irrep of the velocity vector components ($v_i$).
The corresponding irrep-dependent vibrational density of states is
\begin{equation}
\tilde{C}^{v,\Gamma}_{\alpha} (\omega)\propto \len{\int_{-\infty}^{\infty} {\rm d}t C^{v,\Gamma}_{\alpha} (t)e^{-i\omega t}}^2. 
\label{eq:FT_v}
\end{equation}
The site symmetry of Sn is $O_h$, such that the irreps for $(x,y,z)$ are $T_{1u}$.
The site symmetry of Br is $D_{4h}$. 
We define the primary rotation axis (redefined as the $z$-direction) as the unit vector along the direction of the average Br-Sn bond vector in the averaged cubic crystal structure. 
The $D_{4h}$ site symmetry of Br leads to two irreps for vibrations: $E_u$ for $(x,y)$ vibrations and $A_{2u}$ for $z$.
The irrep-dependent vibrational densities of states uncover the symmetry of the vibrations that contribute to the various peaks in the total and atomic vibrational densities of states.
For Br, fluctuations within the $A_{2u}$ irrep result in the peaks near 40, 90, 130, and 170~cm$^{-1}$.
These fluctuations are along the local $z$-axis, oriented along the Br-Sn direction. 
As a result, the $A_{2u}$ fluctuations of Br should couple to vibrations of Sn.
Indeed, the $T_{1u}$ vibrations of Sn exhibit peaks at the same frequencies as the Br $A_{2u}$ density of states.
Moreover, these conclusions agree with previous assignments of Sn-Br stretching modes at frequencies above 100~cm$^{-1}$
and distortions in the Sn-Br$_6$ octahedron below 100~cm$^{-1}$~\cite{yaffe2017local,gao2021metal}.
In contrast, Br vibrations within the $E_{u}$ irrep are low frequency, less than 100~cm$^{-1}$, suggesting that displacements within the local $xy$-plane occur at the same frequencies as vibrations involving Cs.
Based on previous assignments, we expect (and discuss more later) that these low frequency modes involve octahedral tilting~\cite{yaffe2017local,gao2021metal}.
We will discuss vibrational fluctuations involving Cs later, and focus mainly on fluctuations involving Br and Sn in what follows.
%

 \begin{figure}
{\includegraphics[width=0.5\textwidth]{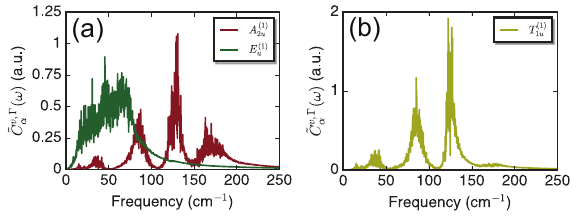}}
  \caption{(a) Symmetry-dependent Br vibrational density of states for the $A_{2u}$ irrep (along local $z$-axis) and $E_u^{(1)}$ irrep (along local $xy$-plane). 
  (b) Symmetry-dependent Sn vibrational density of states for $T_{1u}$ irrep.}
  \label{fig:br_sn_vacf}
\end{figure}

\subsection{Local Electronic Fluctuations are Coupled to Nuclear Fluctuations}

\begin{figure}
{\includegraphics[width=0.48\textwidth]{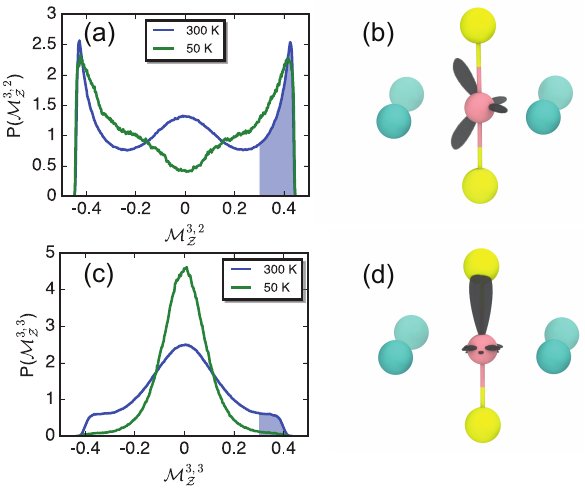}}
  \caption{(a) Probability distribution function (PDF) of the Br-MLWFC $\Mb_{\mathcal{Z}}^{3,2}$ rotor function at 300~K and 50~K.
  (b) Constrained average orientational probability distribution, $f_{\mu,c}(\Theta)$, for Br MLWFCs, where the average was taken over the shaded region in panel a. Also shown is a configuration of Br, Sn, and Cs ions in the perfect cubic crystal structure for comparison. 
  (c) Probability distribution function (PDF) of the Br-MLWFC $\Mb_{\mathcal{Z}}^{3,3}$ rotor function at 300~K and 50~K. 
  (d) Constrained average orientational probability distribution, $f_{\mu,c}(\Theta)$, for Br MLWFCs, where the average was taken over the shaded region in panel c. Also shown is a configuration of Br, Sn, and Cs ions in the perfect cubic crystal structure for comparison.}
  \label{fig:odf_cut}
\end{figure}

The Br-MLWFC unit is a distorted tetrahedron, such that the instantaneous symmetry is incompatible with the ensemble-averaged symmetry of $D_{4h}$.
Instead, the instantaneous symmetry of the unit composed of Br and its four MLWFCs depends on the specific configuration of the nuclei surrounding it. 
Nuclear fluctuations lead to fluctuations in the orientation of the Br-MLWFC unit that can lower the local electronic symmetry, consistent with subduction to a subgroup of $D_{4h}$.
We quantify these symmetry fluctuations through the probability distribution function (PDF) of specific rotor functions,
\begin{equation}
P(\Mb_{\xi}^{\ell,m}) = \avg{\delta\para{\Mb_{\xi}^{\ell,m}-\Mb_{\xi}^{\ell,m}(\Rbar)}}, 
\end{equation}
where $\Mb_{\xi}^{\ell,m}(\Rbar)$ is the value of the rotor function computed in the instantaneous configuration $\Rbar$.
The probability distribution $P(\Mb_{\mathcal Z}^{3,2})$ exhibits are range of fluctuations, Fig.~\ref{fig:odf_cut}a.
These fluctuations suggest breaking of the local symmetry due to structural fluctuations. 
Moreover, $P(\Mb_{\mathcal Z}^{3,2})$ has global maxima at $\Mb_{\mathcal Z}^{3,2}\approx\pm0.4$, indicating that the most probable structures have symmetries within the $E_u$ irrep, 
These Br-MLWFC orientations are similar to those in Fig.~\ref{fig:rotor_examples}b.
At 300~K, we find a local maximum at $\Mb_{\mathcal Z}^{3,2}=0$,
which becomes the minimum of the distribution when the temperature is reduced to 50~K.
We find a similar range of fluctuations in $P(\Mb_{\mathcal Z}^{3,3})$, but this distribution is peaked about zero, Fig.~\ref{fig:odf_cut}c. 
However, there are significant fluctuations away from zero, including small plateaus near $\Mb_{\mathcal Z}^{3,3}=\pm0.4$,
which correspond to electronic structures within the $A_{2u}$ irrep, Fig.~\ref{fig:rotor_examples}a.
As the system is cooled to 50~K, the plateaus near $\Mb_{\mathcal Z}^{3,3}=\pm0.4$ disappear and the distribution becomes more sharply peaked about 0. 
This, combined with the increased peaks at $\Mb_{\mathcal Z}^{3,2}\pm0.4$, suggests that CsSnBr$_3$ has a distorted structure at 50~K that orients Br-MLWFCs with a symmetry within the $E_u$ irrep.
To further illustrate how structural distortions alter the local symmetry of the electron density around Br, we compute the orientational distribution function averaged over configurations within a specified range of $\Mb_{\xi}^{\ell,m}$ values, indicated by the shaded regions in Fig.~\ref{fig:odf_cut}a,c.
This constrained orientational distribution is given by 
\begin{align}
f_{\mu, \rm c}\para{\Theta}&=\frac{1}{4\pi} + \nonumber \\
&\sum_{\ell=1}^{\infty}\ \sum_{m=0}^{2\ell} \frac{\avg{H\para{\Mb_{\xi}^{\ell,m}(\Rbar;\mu)-\Mb_{\rm cut}} \Mb_{\xi}^{\ell,m}}\mathcal{S}_{\xi}^{\ell,m}} {\avg{H\para{\Mb_{\xi}^{\ell,m}(\Rbar;\mu)-\Mb_{\rm cut}}}},
\end{align}
where $H(x)$ is the Heaviside step function and $\Mb_{\rm cut}$ is a cutoff defining the range of values to be considered in the averaging of the rotor function.
Because the distributions are symmetric about zero, we only consider constraining the rotor values over ranges of positive values. 
In comparison to the total average orientational distribution function, $f_\mu(\Theta)$ (Fig.~\ref{fig:odf}a), $f_{\mu, \rm c}(\Theta)$ have a lower symmetry, and the symmetry of each $f_{\mu, \rm c}(\Theta)$
is consistent with a subgroup of $D_{4h}$. 
Constraining $\Mb_{\mathcal Z}^{3,2}$ to large positive values results in a $f_{\mu, \rm c}(\Theta)$ with $C_{2v}$ symmetry, Fig.~\ref{fig:odf_cut}b.
The shape of this distribution results from two Br-MLWFCs each pointing toward a neighboring Sn when the coordination structure distorts away from the ideal cubic structure, while the other two MLWFCs point away from the neighboring Sn ions. 
When constraining $\Mb_{\mathcal Z}^{3,3}$ to large positive values, $f_{\mu, \rm c}(\Theta)$ has $C_{4v}$ symmetry, Fig.~\ref{fig:odf_cut}d. 
The shape of this distribution is consistent with polarization along the Br-Sn axis, suggesting that fluctuations of $\Mb_{\mathcal Z}^{3,3}$ away from zero results from nuclear fluctuations along the local $z$-axis.  
%

\begin{figure}
     {\includegraphics[width=0.48\textwidth]{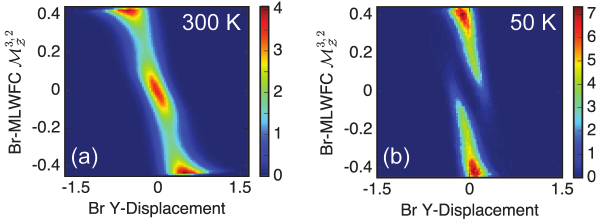}}  
	\caption{Joint probability distributions of the Br-MLWFC rotor function $\Mb^{3,2}_{\mathcal{Z}}$ ($E_u$ symmetry) and the displacement of the Br atom in the $y$-direction at (a) 300~K and (b) 50~K. }
	\label{fig:2dp}
\end{figure}

%
For Br and its four MLWFCs, $E_{u}$ and $A_{2u}$ are the irreps of the displacement vectors in the $xy$-plane and along the $z$-axis, respectively.
To quantify correlations between fluctuations of the Br electronic symmetry and its displacements along these directions,
we define the two-dimensional probability distributions
\begin{equation}
P(\Mb_{\xi}^{\ell,m},\Delta d) = \avg{\delta\para{\Mb_{\xi}^{\ell,m} - \Mb_{\xi}^{\ell,m}(\Rbar)}\delta\para{\Delta d - \Delta d(\Rbar)}},
\end{equation}
where $\Delta d(\Rbar) = d(\Rbar)-d_{\rm eq}$ is the deviation of the Br atom in configuration $\Rbar$ along a particular local axis ($d=x,y,z$) from its equilibrium position $d_{\rm eq}=\avg{d(\Rbar)}$.
We find clear correlations between bromide displacements and Br-MLWFC symmetry, evidenced by
the $P(\Mb_{\mathcal Z}^{3,2},\Delta y)$ distributions shown in Fig.~\ref{fig:2dp}.
The choices of $l$ and $m$ are made so that the rotor function and the displacement are members of the same $E_u$ irrep.
We note that these choices are not unique and other combinations of $l$ and $m$ may yield qualitatively similar behaviors to those identified here, and we use the above choices as illustrative examples of the correlations between local electronic symmetry and nuclear displacements.
At 300~K, $P(\Mb_{\mathcal Z}^{3,2},\Delta y)$ displays three peaks, one near $(0,0)$, and two near approximately $(\pm0.4,\mp0.5)$. 
When the rotor function is zero, the symmetry of the Br-MLWFC unit is orthogonal to that of the rotor function. 
Structurally, these configurations correspond to those where the bromide octahedra are not bending along the local $y$-axis, which is discussed further below. 
The other two peaks, located at non-zero rotor function and non-zero displacements correspond to Br-MLWFC orientations within the $E_u$ irrep when the Br are displaced beyond the equilibrium position.
These values of the rotor function and displacement arise from configurations with finite Sn-Br-Sn bending, which is consistent with tilted octahedra. 
At 50~K, there are only two asymmetric peaks; the peak at $(0,0)$ is not found. 
The absence of the $(0,0)$ peak in the distribution indicates that the bromide ions are distorted away from the averaged cubic unit cell structure throughout the duration of the simulation, suggesting the existence of a low temperature phase with frozen-in local disorder.
Such a disordered low temperature phase was recently identified~\cite{fabini2024noncollinear}.

\subsection{Local Electronic Fluctuations are Dynamic}

The correlations between nuclear displacements and the orientation of local electronic structure are not static.
The nuclei, and consequently the local electronic symmetry, fluctuate dynamically. 
To quantify these dynamic electronic fluctuations, we computed time correlation functions of appropriate rotor functions
characterizing the orientation of Br-MLWFC and Sn-MLWFC units.
We computed the rotor function TCF as
\begin{equation}
C^{\ell, \Gamma^{(i)}}_{\mu}(t) = \frac{\sum\limits_{m} \delta_{\Gamma^{(i)},\tilde{\Gamma}(\ell,m)}\left\langle{\mathcal{M}}_{\xi}^{\ell, m}(0;\mu) {\mathcal{M}}_{\xi}^{\ell, m}(t;\mu) \right\rangle}{\sum\limits_{m} \delta_{\Gamma^{(i)},\tilde{\Gamma}(\ell,m)}\left\langle{(\mathcal{M}}_{\xi}^{\ell, m}(0;\mu))^2 \right\rangle} ,
\label{eq:cor}
\end{equation}
where $\Gamma^{(i)}$ is the irrep label ($\Gamma$) of the $i$th symmetrized basis of that label, $\tilde{\Gamma}(\ell,m)$ is a function that returns the irrep label, $\Gamma^{(j)}$, corresponding to the specific combination of $\ell$ and $m$, and $\delta_{\alpha,\beta}$ is the Kronecker delta function.
To identify the frequencies of electronic structure fluctuations, we also computed the power spectra of the normalized TCFs,
\begin{equation}
\tilde{C}^{\ell,\Gamma^{(i)}}_{\mu}(\omega) \propto \int_{-\infty}^{\infty}  C^{\ell,\Gamma^{(i)}}_{\mu}(t)e^{-i\omega t}{\mathrm{d}t}.
\label{eq:FT}
\end{equation}

\begin{figure}
{\includegraphics[width=0.5\textwidth]{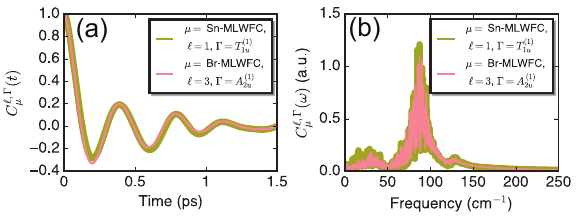}} 
  \caption{(a) Time correlation function and (b) power spectrum of the Br-MLWFC and Sn-MLWFC rotor functions.}
  \label{fig:MLWFC_PS}
\end{figure}

The symmetry-dependent vibrational densities of states shown in Fig.~\ref{fig:br_sn_vacf} suggest that Br ions exhibit
fast, high frequency $A_{2u}^{(1)}$ fluctuations and slow, low frequency $E_u^{(1)}$ fluctuations.
Consequently, the electronic symmetry fluctuations that couple to vibrations should also exhibit two symmetry-dependent timescales. 
For the Br-MLWFC fluctuations, we do find fast and slow decays for different irreps, and we focus on the faster electronic fluctuations here before discussing the slower modes in Section~IV.D.
The high frequency $A_{2u}^{(1)}$ vibrations of Br are similar to the high frequency $T_{1u}^{(1)}$ vibrations of Sn.
As a result, we can also expect the fast timescale for electronic fluctuations of Br to be similar to those for Sn.
Indeed, we find that the Br-MLWFC fluctuations of $A_{2u}^{(1)}$ behave nearly identical to those for Sn-MLWFC of $T_{1u}^{(1)}$, both of which quantify fluctuations in the local $z$-direction.
The time correlation functions for these two types of fluctuations decay and oscillate on the same timescales, Fig.~\ref{fig:MLWFC_PS}a.
Their power spectra overlap and exhibit a large peak near 90~cm$^{-1}$ and two minor peaks near 40~cm$^{-1}$ and 140~cm$^{-1}$,
Fig.~\ref{fig:MLWFC_PS}b, further suggesting
that Br-MLWFC and Sn-MLWFC fluctuations are dynamically coupled. 
%

%
To explicitly quantify time-dependent correlations between Br-MLWFC and Sn-MLWFC orientations,
we compute the time-dependent cross-correlation function,
\begin{equation}
C^{\zeta, \zeta^{\prime}}_{\mu\mu^{\prime}}(t) = \avg{\zeta(0;\mu) \zeta^{\prime}(t;\mu^{\prime})},
\end{equation}
where $\zeta$ indicates the type of function being computed, such as a rotor function ($\zeta=\Mb_{\xi}^{\ell,m}$) or displacement ($\zeta=\Delta d$).
The cross-correlation function, $C^{\zeta, \zeta^{\prime}}_{\mu\mu^{\prime}}(t)$, can be non-zero if $\zeta$ and $\zeta'$ are partner functions to the same irrep and undergo subduction to the same point group and primary axis.
For example, within the point group $D_{4h}$, we can consider displacements along the local $x$-axis and $\mathcal{M}^{1,0}_{\mathcal{Z}}$, both of which are partner functions of $E_u$.
However, displacements along the $x$-axis undergo subduction to C$_{2v}$ with the primary axis equal to the $x$-axis, while $\mathcal{M}^{1,0}_{\mathcal{Z}}$ undergoes subduction to C$_{2v}$ with the primary axis equal to the $y$-axis.
As a result, the corresponding $C^{\zeta, \zeta^{\prime}}_{\mu\mu^{\prime}}(t) = 0$.
Because of this condition on the cross-correlations between $\zeta$ and $\zeta'$,
if we want to quantify correlations between MLWFC orientations on two sites of different symmetry, or between MLWFC orientations and atomic displacements, 
we need to construct a collective rotor function that is a partner function of the appropriate irrep and shares a primary axis with the displacement mode or another rotor function of a site with a different symmetry (e.g. $O_h$ for Sn or $D_{4h}$ for Br).
To enable the calculation of cross-correlation functions,
we define a collective rotor function, $\tilde{\mathcal{M}}^{(\ell,m),\Gamma}_{\xi}$, as a linear combination of the $\mathcal{M}^{\ell,m}_{\xi}$ for the appropriate nearest neighbors of a central atom using a projection operator formalism (see Section S5 for additional details).
This collective rotor function can have a non-zero cross-correlation with a $\zeta$ of irrep $\Gamma$.

\begin{figure}
   {\includegraphics[width=0.48\textwidth]{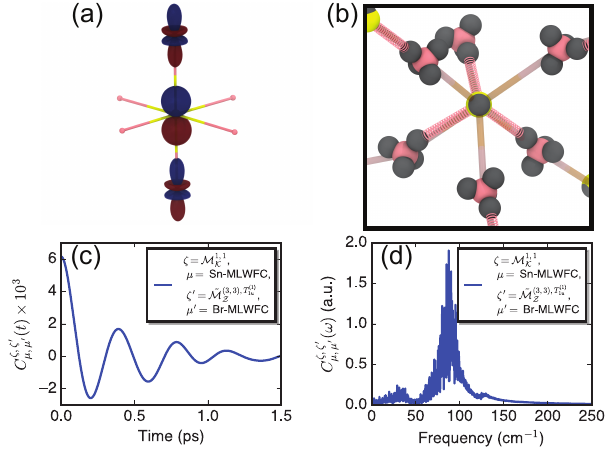}}
     \caption{(a) Illustration of the collective rotor function that is a partner function to $T_{1u}$ with respect to Sn, which consists of the sum of $\mathcal{M}^{3,3}_{\mathcal{Z}}$ rotor functions for Br-MLWFC units along the same axis, appropriately rotated, resulting in a function with $C_{4v}$ symmetry. Shown are the two $\mathcal{M}^{3,3}_{\mathcal{Z}}$ centered on Br ions, as well as the $\Mb^{1,1}_{\mathcal{K}}$ rotor function centered on the Sn ion at the center.
     (b) Simulation snapshot showing the coupling between off-centering of the Sn-MLWFC unit ($T_{1u}$) and the Br-MLWFC orientation ($A_{2u}$). Sn are yellow, Br are pink, and MLWFCs are gray. Directional orientation of Br MLWFCs toward Sn are highlighted by pink dashed cylinders.
     (c) Time-dependent cross-correlation function between the Sn-MLWFC rotor function and the Br-MLWFC collective rotor function,
     and (d) the corresponding power spectrum.
    }
  \label{fig:collective}
\end{figure}

Coupling between Br-MLWFC and Sn-MLWFC orientations causes the halide to experience out-of-phase vibrations, breaking the symmetry from $D_{4h}$ to $C_{4v}$.
For Sn, its environment can break from $O_h$ to $C_{3v}$ or $C_{4v}$, depending on the number out-of-plane vibrations that are coupled.
One of the motions belonging to this irrep is associated with off-centering of the B-site cation along the [111] direction~\cite{gao2021metal}, which results in $C_{3v}$ symmetry. 
Therefore, we quantify cross-correlations between the Sn-MLWFC rotor function of the $T_{1u}$ irrep and a collective rotor function of the Br-MLWFC units. 
We construct this collective rotor function by summing the $\Mb_{\Zb}^{3,3}$ for the Br-MLWFC along the same axis, appropriately rotated, Fig.~\ref{fig:collective}a.
The resulting cross-correlation function and the corresponding power spectrum, Fig.~\ref{fig:collective}c,d, closely mirror the rotor function time correlation functions for Br-MLWFC and Sn-MLWFC, indicating strong correlation between the local electronic symmetries of Br and Sn.
The coupling of these symmetries is a consequence of the preferential orientation of the Sn lone pair when it off-centers,
and further suggests that off-centering of B-site ions containing a lone pair is a collective process that requires reorientation of both the Sn-lone pair unit and the electron pairs of the neighboring Br ions, as illustrated by the snapshot in Fig.~\ref{fig:collective}b.
%


\subsection{Local Electronic Symmetry is Coupled to Octahedral Tilting}

\begin{figure}
	\centering
        \includegraphics[width = 0.48\textwidth]{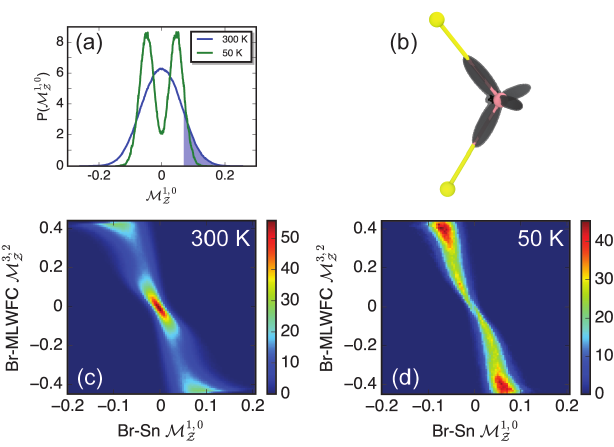}
	\caption{Correlations between Sn-Br-Sn bending and the Br-MLWFC orientation ($E_u$) quantified by the joint probability distribution of their respective rotor functions at (a) 300~K and (b) 50~K.
	(c) The Sn-Br-Sn bending histogram at 300~K and 50~K. 
	(d) The constrained average orientational distribution function, $f_{\mu,c}(\Theta)$, where the rotor function is constrained to lie in the shaded region of the 300~K Sn-Br-Sn bending histogram. The distribution function is shown as a gray isosurface, and a representative configuration of a Sn-Br-Sn triplet is shown.}
	\label{fig:br-sn}
\end{figure}

\begin{figure}
   {\includegraphics[width=0.4\textwidth]{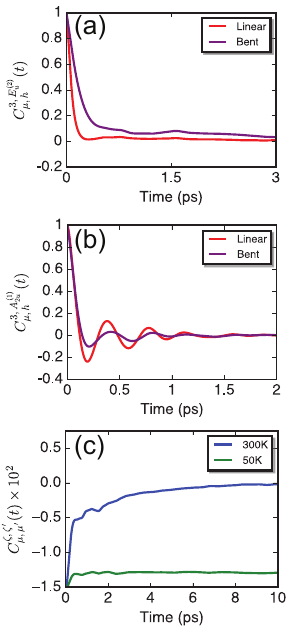}}
  \caption{(a,b) The time correlation functions for Br-MLWFC (a) $E_u^{(2)}$ and (b) $A_{2u}$ rotations when the octahedron is bent ($\len{\Mb_{\mathcal Z}^{1,0}}>0.07$) or not bent ($\len{\Mb_{\mathcal Z}^{1,0}})<0.025$).
   (c) Time-dependent cross-correlation function between the Br-MLWFC rotor function and the Sn-Br-Sn rotor function.}
  \label{fig:tilt-cors}
\end{figure}

To illustrate the coupling between electronic symmetry fluctuations and specific phonon modes, we consider the changes in local electronic symmetry induced by octahedral tilting. 
In metal halide perovskites, the halide-metal octahedra do not remain in their average cubic structure.
Instead, the softness of the lattice allows the octahedra to tilt out of alignment in pairs, with significant consequences on many of their material and photophysical properties~\cite{beecher2016direct,yang2017spontaneous,fabini2016dynamic,laurita2017chemical,zhao2020polymorphous,Maity:2022aa,fabini2020,mozur2021cation,laurita2022,wiktor2023quantifying}. 
We first need an order parameter to quantify octahedral tilting, before connecting tilting to changes in electronic structure.
The perovskite lattice can be described as vertex-sharing B-X octahedra,
such that the dynamics of the octahedra can be looked at from the perspective of the B-site cation or the X-site anion. 
From the perspective of the Br, we can define a rotor function that quantifies Sn-Br-Sn bending, which is coupled to octahedral tilting (though bending does not necessarily always imply tilting). 
Sn-Br-Sn bending transforms as $E_u$, breaking the symmetry from $D_{4h}$ to C$_{2v}$, and we quantify this bending with $\Mb_{\mathcal Z}^{1,0}$ for the Br-Sn vectors. 
The probability distribution of $\Mb_{\mathcal Z}^{1,0}$ indicates significant Sn-Br-Sn bending at 300~K, Fig.~\ref{fig:br-sn}a.
While the distribution is peaked at zero, corresponding to no bending, there are significant deviations away this value,
indicating dynamic Sn-Br-Sn bending. 
Interestingly, at 50~K, the distribution splits into two peaks with a minimum at zero. 
The two peaks correspond to Sn-Br-Sn bending in two directions, and the split suggests that bent structures
are stable at this low temperature.
Because the Sn-Br-Sn angle is altered by octahedral tilting, this further suggests that the octahedra rarely fluctuate between tilted configurations at 50~K. 
This is consistent with recent results that identified the structure of low temperature phases of CsSnBr$_3$~\cite{fabini2024noncollinear}.
Our results above for the constrained orientational distribution, $f_{\mu, \rm c}(\Theta)$, suggest that
the symmetry of the local electron density couples to distortions of the Br-Sn coordination environment.
Therefore, Sn-Br-Sn bending should induce similar changes in the local symmetry of the Br-MLWFC unit. 
We computed the Br-X $f_{\mu, \rm c}(\Theta)$ while constraining the Br-Sn $\Mb_{\mathcal Z}^{1,0}>0.07$, Fig.~\ref{fig:br-sn}b.
The Br-X $f_{\mu, \rm c}(\Theta)$ induced by octahedral tilting is consistent with $C_{2v}$ symmetry, significantly different from the average $D_{4h}$ symmetry, which indicates a significant coupling between the local Br electronic density and Sn-Br-Sn bending. 
The coupling between Br-MLWFC orientation and Sn-Br-Sn bending can be further quantified through joint probability distributions of the Br-MLWFC $\Mb_{\Zb}^{3,2}$ and the Br-Sn $\Mb_\Zb^{1,0}$ rotor functions, Fig.~\ref{fig:br-sn}c,d.
We find features reminiscent of the joint distributions for Br-MLWFC orientation and Br atom displacements.
We find a peak at (0,0) that results from linear Sn-Br-Sn configurations and $A_{2u}$ symmetry of the Br-MLWFC unit, Fig.~\ref{fig:br-sn}c.
We also find peaks at approximately $(\mp0.1,\pm0.4)$, which result from bent Sn-Br-Sn configurations and the Br-MLWFC unit orienting two electron pairs to the nearest Sn ions, consistent with the $f_{\mu,\rm c}(\Theta)$ shown in Fig.~\ref{fig:br-sn}b.
When the system is cooled to 50~K, the peak at (0,0) once again disappears while the peaks at $(\mp0.1,\pm0.4)$ remain, consistent with tilted octahedra persisting throughout the duration of the simulation.
We find that there is dynamic coupling between the local symmetry of the Br-MLWFC unit and octahedral tilting by computing a modified version of the total rotor function TCF,
\begin{equation} C^{\ell,\Gamma^{(i)}}_{\mu,h}(t) = 
\frac{\sum_{m} \delta_{\Gamma^{(i)},\tilde{\Gamma}(\ell,m)} \avg{h(0) \Mb_{\xi}^{\ell,m}(0;\mu) h(t) \Mb_{\xi}^{\ell,m}(t;\mu)}}{\sum_{m} \delta_{\Gamma^{(i)},\tilde{\Gamma}(\ell,m)} \avg{(h(0) \Mb_{\xi}^{\ell,m}(0;\mu))^2} },
\end{equation}
where $h(t)$ is an indicator function.
To probe electronic fluctuations when the octahedra are tilted, $h(t)=1$ if the Br-Sn $\len{\Mb_{\mathcal Z}^{1,0}}>0.07$ and $h(t)=0$ otherwise. 
Similarly, to quantify dynamic symmetry fluctuations when the octahedra are not tilted, $h(t)=1$ if the Br-Sn $\len{\Mb_{\mathcal Z}^{1,0}}<0.07$ and $h(t)=0$ otherwise. 
The resulting $C^{3,E_u^{(2)}}_{\mu,h}(t)$ indicate that the MLWFC symmetry changes faster when the Sn-Br-Sn unit is linear, and the symmetry fluctuations are slower when the Sn-Br-Sn unit is bent, Fig.~\ref{fig:tilt-cors}a, strongly suggesting a coupling between octahedral tilting and localized electronic structure and dynamics.
We can use the same kind of analysis to examine the response of the $A_{2u}$ orientational fluctuations to Sn-Br-Sn bending. 
The resulting $C^{3,A_{2u}^{(1)}}_{\mu,h}(t)$ indicates that the $A_{2u}$ orientational fluctuations are damped by Sn-Br-Sn bending as compared to when the unit is linear, Fig.~\ref{fig:tilt-cors}b.
This may be expected from the favorable interactions between the two Br-MLWFCs with two Sn ions, which stabilize the bent configuration.
To further quantify the coupling between local electronic symmetry and Sn-Br-Sn bending,
we computed time-dependent cross-correlation functions .
When $\zeta = \Mb_{\mathcal Z}^{3,2}$, $\mu=$(Br, MLWFC) and $\zeta^{\prime} = \Mb_{\mathcal Z}^{1,0}$, $\mu'=$(Br, Sn), we find that the cross-correlation is negative, indicating an anticorrelation between the orientation of the Br-MLWFC unit and bending of the Sn-Br-Sn unit, Fig.~\ref{fig:tilt-cors}c.
Note that both rotor functions are of the same irrep, so we do not need to define a collective rotor function to probe cross-correlations involving orientational fluctuations. 
The cross-correlation, $C^{\zeta, \zeta^{\prime}}_{\mu\mu^{\prime}}(t)$, decays somewhat smoothly to zero on a timescale of roughly 5~ps, consistent with timescales relevant to octahedral tilting~\cite{gao2021metal,liang2023structural,lanigan2021two,weadock2023nature} and the above-discussed fluctuations in local electronic symmetry. 
The decay is not described completely by a single exponential, and multiple timescales are involved in the decay, consistent with the multiple peaks found in the vibrational density of states. 
When the system is cooled to 50~K, $C^{\zeta, \zeta^{\prime}}_{\mu\mu^{\prime}}(t)$ starts at a similar negative value at $t=0$, but only exhibits a rapid, short-time decay before plateauing near $-0.013$. 
This lack of decay is consistent with the octahedra remaining tilted throughout the duration of the simulation
and a negligible fraction of non-tilted configurations. 
%

\begin{figure}
 {\includegraphics[width=0.47\textwidth]{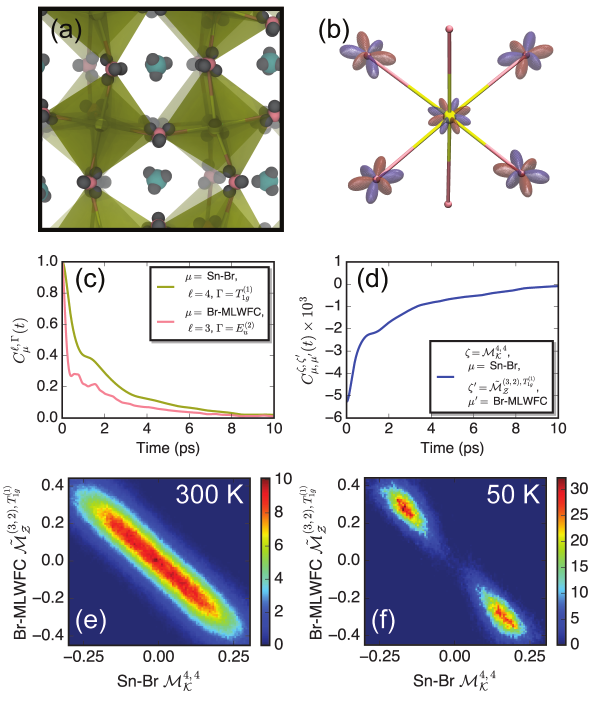}}
  \caption{(a) Representative snapshot of octahedral tilting, illustrating the coupling between tilt and Br-MLWFC orientation. 
  Sn are yellow, Br are pink, Cs are cyan, and MLWFCs are yellow. Note the alignment of the Br MLWFCs with the tilt angle.
  (b) Illustration of the collective rotor function that is a partner function to $T_{1g}$, which consists of the sum of $\mathcal{M}^{3,2}_{\mathcal{Z}}$ rotor functions for in-plane Br-MLWFC units, rotated appropriately, resulting in a function that has $C_{4h}$ symmetry.
     (c) Time-dependent autocorrelation functions of Sn-Br octahedral tilting ($T_{1g}$) and Br-MLWFC orientation ($E_u^{(2)}$). 
  (d) Time-dependent cross correlation function between Sn-Br ($T_{1g}$) and the Br-MLWFC collective rotor function ($E_u^{(2)}$).
  (e,f) Joint probability distribution of the Br-MLWFC and Sn-Br rotor functions at (e) 300~K and (f) 50~K. }
    \label{fig:collective-tilt}
\end{figure}

%
Octahedral tilting arises when four Br within the same plane and octahedron bend along the same rotation axis.
This collective fluctuation can be studied using a rotor function dependent on the Sn-Br octahedra.
Octahedral tilting of the B-site cation transforms as the irrep $T_{1g}$, breaking the symmetry from $O_h$ to C$_{4h}$. 
This dynamic local symmetry breaking results in a propensity for Br-MLWFCs to orient with the tilting of the octahedra, evident in the simulation snapshot in Fig.~\ref{fig:collective-tilt}a.
Therefore, we computed the time correlation function of the Sn-Br $\Mb_{\mathcal{K}}^{4,4}$ rotor function ($T_{1g}^{(1)}$ irrep)
and compare it to the time correlation function of the Br-MLWFC rotor function ($E_u^{(2)}$), Fig.~\ref{fig:collective-tilt}c.
Both correlation functions decay on a similar, slow timescale of roughly 5~ps, consistent with timescales for dynamic octahedral tilting~\cite{gao2021metal,liang2023structural,lanigan2021two,weadock2023nature}.
The similarity of these correlation functions suggests further coupling between Br-MLWFC orientation and Sn-Br-Sn bending.
%

%
Octahedral rotation involves the collective displacement of atoms and the coupling of this collective displacement to local changes in electronic structure. 
To quantify this collective displacement and its coupling to electronic structure,
we define another collective rotor function that is a partner function of the $T_{1g}$ irrep as the sum of Br-MLWFC $E_{u}$ rotor functions around the Sn, where the sum is over the four in-plane bromides, Fig.~\ref{fig:collective-tilt}b.
Using this collective rotor function, $\tilde{\Mb}_{\mathcal{Z}}^{(3,2),T_{1g}}$, we can compute the time-dependent cross-correlation function of collective Br orientations and Sn-Br bond orientations.
The cross-correlation function indicates anticorrelation and decays on the same slow timescale as individual Sn-Br orientational correlations and the slow Br-MLWFC fluctuations of $E_u^{(2)}$ symmetry, Fig.~\ref{fig:collective-tilt}b, suggesting that the electron orientation is strongly coupled to octahedral rotation.
Indeed, static correlations suggest a nearly linear relationship between the collective Br-MLWFC rotor function and the Sn-Br rotor function, Fig.~\ref{fig:collective-tilt}e, further indicating strong correlation.
This linear correlation persists at 50~K, Fig.~\ref{fig:collective-tilt}f, although the distribution is split into two peaks at non-zero values of the rotor functions due to the static octahedral tilt induced by the low temperature.
%

\begin{figure*}
{\includegraphics[width=0.9\textwidth]{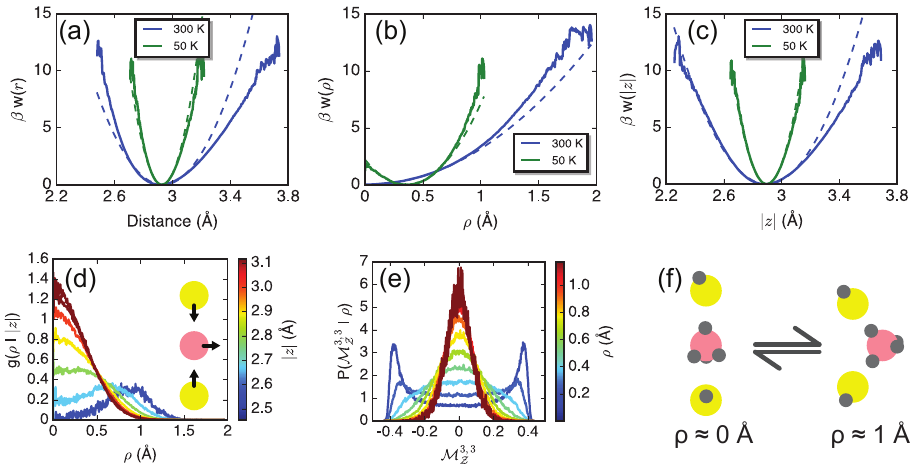}}
  \caption{(a) The Br-Sn potential of mean force, $-\ln g(r)$, in units of $\kT$, where $g(r)$ is the Br-Sn raidal distribution function.
  (b,c) The potential of mean force decomposed into contributions from correlations (b) in the local $xy$-plane and (c) along the local $z$-axis. Dashed lines indicate Gaussian fits performed near the minimum of $-\ln g(r)$.
  (d,e) The conditional radial distribution functions for (d) correlations in the $xy$-plane given a value of $\len{z}$ and (e) correlations along the local $z$-axis given $\sqrt{x^2+y^2}$. Sketches of the vibrational modes that lead to these correlation functions are shown as insets. Both Sn ions (yellow) move along $z$ toward or away from the bromide (pink), which displaces in the $xy$-plane.
  (f) Schematic illustration of the change in MLWFC orientation upon bending of the Sn-Br-Sn unit. Sn ions are drawn in yellow, Br ions in pink, and MLWFCs in gray. Shown are representative configurations when the Sn-Br-Sn triplet is linear, $\rho\approx0$~\AA, and bent, $\rho\approx1$~\AA.}
  \label{fig:rdf}
\end{figure*}

\subsection{Phonon Anharmonicity is Coupled to Electron Density Fluctuations}
Halide perovskites are relatively soft solid materials that display significant anharmonic vibrational fluctuations. 
These anharmonic vibrational fluctuations can be observed in appropriate radial distribution functions~\cite{schilcher2023correlated}, such as the Br-Sn radial distribution function, $g(r)$, Fig.~\ref{fig:rdf}.
At 300~K, the first peak in the Br-Sn radial distribution function, $g(r)$, is significantly non-Gaussian far from the peak.
Consequently, this anharmonic Br-Sn $g(r)$ results in an anharmonic potential of mean force, $\beta w(r)=-\ln g(r)$, quantifying the effective interactions between the two ions.
When the temperature is lowered to 50~K, $w(r)$ becomes approximately harmonic, consistent with the system being trapped in a tilted configuration.
The origins of the pronounced asymmetry in the Br-Sn $g(r)$ and the resulting models for interpreting radial distribution functions have been a matter of debate~\cite{laurita2017chemical,bridges2023local}, and our analysis of local nuclear and electronic structure fluctuations described below should aid in the interpretation of this asymmetry.
Based on our analysis of symmetry-dependent fluctuations above, we can decompose the $w(r)$ into contributions from
within the local $xy$-plane, Fig.~\ref{fig:rdf}b, and along the local $z$-axis, Fig.~\ref{fig:rdf}c.
Fluctuations within the local $xy$-plane are quantified by the coordinate $\rho=\sqrt{x^2+y^2}$ while those along the local $z$-axis are quantified by $\len{z}$.
At both 300~K and 50~K, $w(\rho)$ is approximately harmonic and even sub-harmonic for large displacements,
indicating that the non-Gaussian features in $g(r)$ do not originate in fluctuations along the in-plane direction.
In contrast, $w(\len{z})$ is significantly anharmonic at 300~K for displacements larger than the mean. 
At 50~K, the potential of mean force is again approximately harmonic, in agreement with experimental results that suggest harmonic lattice motion in halide perovskites at low temperatures~\cite{guo2019dynamic}.
Therefore, the anharmonicity of the Br-Sn potential of mean force results from changes in interactions induced by displacements along the local $z$-axis defined by the Br-Sn bond axis in the ideal cubic structure.
Displacements along the $z$-axis and within the $xy$-plane may not be independent in an anharmonic system,
and so we also quantified the coupling between these two types of displacements with conditional radial distribution functions, Fig.~\ref{fig:rdf}d,e.
Correlations in the $xy$-plane are significantly coupled to the displacement along the $z$-axis, evidenced by $g(\rho~|\len{z})$ in Fig.~\ref{fig:rdf}d.
For large Br-Sn distances along the $z$ direction, $\len{z}$, $g(\rho~|\len{z})$ is peaked at zero, indicating that the Sn-Br-Sn triplet is linear and reminiscent of the cubic structure.
As $\len{z}$ decreases, the peak at $\rho=0$ reduces in magnitude and shifts toward non-zero values, approaching $\rho=1$~\AA \ for the smallest $\len{z}$ values probed here.
The peak shift indicates that as the Sn-Br distance in the $z$ direction is reduced, Br displaces outward and the Sn-Br-Sn triplet bends.
The presence of only a single peak in $g(\rho~|\len{z})$ indicates that both Sn ions move toward or away from the Br ion at a similar distance; otherwise, the peak would be split into two. 
This coupled vibrational motion is schematically indicated by the sketch in the inset of Fig.~\ref{fig:rdf}d.
The conditional radial distribution function for the $z$-direction, $g(\len{z} | \rho)$, is consistent with the same qualitative picture, Fig.~S4.
At small $\rho$, $g(\len{z}|\rho)$ is peaked at larger values of $\len{z}$.
As $\rho$ is increased, the peak position shifts to smaller $\len{z}$, consist with compression of the Sn-Br-Sn unit in the $z$-direction when it bends.
Furthermore, the single peak of the distribution again suggests that the two Sn ions of the Sn-Br-Sn unit are nearly equidistant from the Br along $z$.
Our analysis in previous sections indicates that Sn and Br electronic orientations are coupled along the local $z$-axis, and one might anticipate that the anharmonic fluctuations along the local $z$-axis are coupled to changes in local electronic symmetry.
To quantify this coupling, we computed the conditional probability of the $\Mb_{\Zb}^{3,3}$ rotor function that quantifies the orientation of the Br-MLWFC unit for given values of the in-plane displacement of the bromide, $\rho$, Fig.~\ref{fig:rdf}e.
For small $\rho$, the Sn-Br-Sn unit is linear, and $P(\Mb_{\Zb}^{3,3} | \rho)$ is bimodal and peaked near $\pm0.4$.
These values of $\Mb_{\Zb}^{3,3}$ correspond to Br orientating one MLWFC directly toward Sn, as indicated in the sketch in Fig.~\ref{fig:rdf}f.
As $\rho$ is increased and the Sn-Br-Sn unit bends, the two peaks in the distribution converge to a single peak centered at zero,
consistent with a change in symmetry of the Br-MLWFC orientation.
This change is consistent with the Br ion pointing one MLWFC at each Sn, such that two MLWFCs are now involved in interactions with Sn ions (one for each Sn in the Sn-Br-Sn unit), Fig.~\ref{fig:rdf}f.
Therefore, the vibrational mode sketched in Fig.~\ref{fig:rdf}d is accompanied by changes in local electronic orientation.
We suggest that this electronic rotation--nuclear translation coupling leads to the anharmonicity and phonon softening observed in CsSnBr$_3$ and other halide perovskites, similar to the effects of rotation--translation coupling in molecular plastic crystals~\cite{lynden1994translation}.
The conditional pair distribution functions at 50~K further support the idea that electronic rotation--nuclear translation coupling leads to anharmonicity, Fig.~S5.
At 50~K, the Sn-Br potential of mean force is harmonic, which suggests that the coupling between electronic orientation and nuclear translations should be small. 
Indeed, the structure at 50~K is essentially the bent structure illustrated on the right side of Fig.~\ref{fig:rdf}f throughout duration of the simulation. 
As a result, $g(\rho | \len{z})$ is peaked at finite $\rho$ for all sampled $\len{z}$ values. 
Similarly, $P(\Mb_{\Zb}^{3,3} | \rho)$ is peaked at zero for nearly all configurations and never bimodal, with rare, small-$\rho$ configurations exhibiting a flat distribution that is consistent with equal probability of the two orientations sketched in Fig.~\ref{fig:rdf}f in this small fraction of configurations.
Therefore, because electronic rotations are absent at 50~K, the resulting electronic rotation--nuclear translation coupling is small and the Sn-Br potential of mean force is harmonic.
%

\subsection{Local Electronic Fluctuations of the $A$-site}
\begin{figure}
	\centering
        \includegraphics[width = 0.48\textwidth]{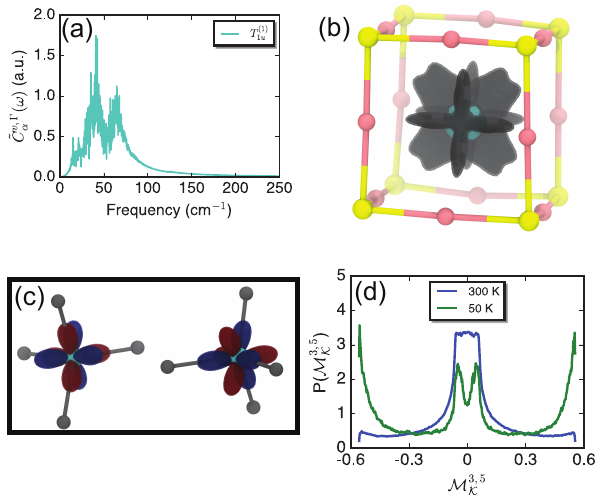}
	\caption{(a) Vibrational density of states for Cs modes of $T_{1u}^{(1)}$ symmetry.
	(b) Orientational distribution function of the Cs-MLWFCs shown within the cubic coordination environment of the Cs. 
	(c) Cs-MLWFC cubic harmonic $\mathcal{S}^{3,5}_{\mathcal{K}}$ ($T_{2u}$) used to quantify the orientations of the Cs-MLWFC units.Two configurations of the Cs-MLWFC with opposite signs are shown.
	(d) Probability distribution function of the Cs-MLWFC rotor function, $\Mb_{\mathcal{K}}^{3,5}$, at temperatures of 300~K and 50~K.}
	\label{fig:cs}
\end{figure}

The precise importance of the $A$-site cation in determining the properties of halide perovskites is unclear~\cite{lee2022}.
The Cs$^+$ electronic states are deep in the valence band~\cite{huang2013}, and as a result, it is usually considered to have negligible impact on the properties of halide perovskites beyond tuning stability.
However, it has been suggested that even a simple $A$-site cation like Cs$^+$ can have a significant impact on the phonon modes of halide perovskites, and consequently an indirect impact on materials properties~\cite{yaffe2017local,zhu2022probing}.
In this section, we demonstrate that the Cs$^+$ cation also exhibits non-trivial MLWFC fluctuations that are coupled to lattice dynamics.
The contribution of Cs to the vibrational density of states, Fig.~\ref{fig:total_vdos}, primarily exhibits low frequency modes below 100~cm$^{-1}$, and we uncover a similar picture through the symmetry-dependent vibrational density of states for the $T_{1u}^{(1)}$ irrep, Fig.~\ref{fig:cs}a.
The vibrational frequencies of Cs overlap significantly with the $E_u$ vibrational density of states for Br, suggesting that the vibrational modes are coupled. 
Indeed, displacement of Cs ions typically consist of Cs moving toward a face of the surrounding Br-Sn cube, resulting in a displacement of the four bromides on the edge of that face, suggesting a coupling between Cs and Br vibrational modes~\cite{yaffe2017local}.
%

 \begin{figure}
   {\includegraphics[width=0.48\textwidth]{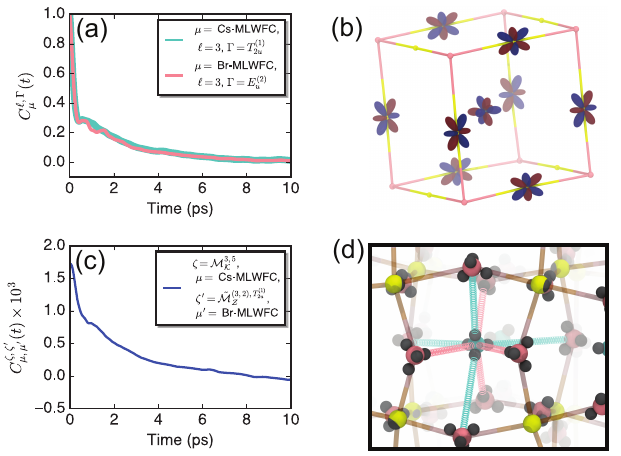}}   
  \caption{(a) Time correlation functions of the Cs-MLWFC orientation ($T_{2u}$) and the Br-MLWFC orientation ($E_u^{(2)}$) rotor functions. 
  (b) The collective rotor function that is a partner function to $T_{2u}$ constructed using the eight nearest neighbor bromides around a central Cs, which consists of the sum of $\mathcal{M}^{3,2}_{\mathcal{Z}}$ on each Br, rotated appropriately, resulting function with $D_{2d}$ symmetry.
  (c) Time-dependent cross-correlation function between the Cs-MLWFC orientation ($T_{2u}$) and the collective rotor function for the Br-MLWFC orientations ($E_u^{(2)}$).
  (d) Simulation snapshot illustrating the coupling of Cs-MLWFC orientation and Br-MLWFC. Dashed cylinders highlight direction preferences of the Cs-MLWFCs toward Br (cyan) and Br-MLWFCs toward Cs (pink).}
    \label{fig:collective_rotor_cs}
\end{figure}

%
The Cs$^+$ cation has four electron pairs, and we can quantify any coupling of their orientation to vibrational modes through the type of analysis used above for Br and Sn. 
The orientational distribution function, $f_\mu(\Theta)$, for the Cs MLWFCs is not spherical, as one might expect, but instead shows significant structure, Fig.~\ref{fig:cs}b.
The MLWFCs preferentially orient toward bromide and away from Sn. 
The non-trivial structure of the Cs orientational distribution function suggests the existence of local Cs electronic symmetry fluctuations that couple to distortions of the surrounding Sn and Br ions. 
To quantify the orientation of the Cs-MLWFC unit, we use rotor functions based on the $\mathcal{S}_{\mathcal{K}}^{3,5}$ cubic harmonic, shown in Fig.~\ref{fig:cs}c, which are partner functions to the $T_{2u}$ irrep. 
The probability distribution of $\Mb_{\mathcal{K}}^{3,5}$ at 300~K displays a single plateau-like peak near zero, with significant fluctuations away from zero. 
At 50~K, when the octahedra remain tilted and rarely change orientation, the probability distribution of the rotor function splits into two peaks near zero and large peaks emerge at the positive and negative bounds on $\Mb_{\mathcal{K}}^{3,5}$.
This splitting of the distribution and emergence of large peaks suggest that the Cs MLWFCs adopt preferential orientations at low temperatures in response to the non-zero average octahedral tilt.
We computed the time correlation function of the $\Mb_{\mathcal{K}}^{3,5}$ rotor function ($T_{2u}^{(1)}$ symmetry) to
quantify the dynamic fluctuations of the Cs-MLWFC orientation, and compare this to the time correlation function of the $E_u^{(2)}$ symmetry fluctuations of the Br-MLWFC unit. 
Joint probability distributions suggest that these electronic orientations are coupled and suggest distinct orientations at low temperatures, Fig.~S3.
The Br-MLWFC fluctuations are coupled to the low frequency $E_u^{(1)}$ vibrational modes, Fig.~\ref{fig:br_sn_vacf}.
These time correlation functions overlap almost completely, Fig.~\ref{fig:collective_rotor_cs}a, suggesting that the Cs-MLWFC and Br-MLWFC fluctuations are dynamically coupled and that the Cs-MLWFC unit reorients in response to octahedral tilting. 
Fluctuations in the Cs-MLWFC order parameter, $\Mb_{\mathcal{K}}^{3,5}$, which is a partner function to the $T_{2u}$ irrep, correspond to a subduction in the local symmetry from $O_h$ to $D_{2d}$.
The resulting $D_{2d}$ local symmetry is what one would expect Cs to experience from out-of-phase octahedral tilting ($a^0a^0c^-$ in Glazer's notation), further suggesting a coupling of Cs-MLWFC orientation and octahedral tilting. 
We can quantify the coupling between Cs-MLWFC orientation and the surrounding bromide cage through a time-dependent cross-correlation function.
This time correlation function again involves a collective rotor function, which we define as the sum of $E_u$ functions for the eight Br nearest neighbors of Cs, resulting in a partner function of the $T_{2u}$ irrep, Fig.~\ref{fig:collective_rotor_cs}b.
The cross-correlation function closely tracks the decay of the individual time correlation functions for Cs-MLWFC and Br-MLWFC, suggesting that indeed the Cs-MLWFC and Br-MLWFC orientations are dynamically coupled on the slow timescale or low frequencies typical of octahedral tilting, Fig.~\ref{fig:collective_rotor_cs}c.
This dynamic coupling is a result of preferential orientation of Cs-MLWFCs in response to the orientations Br-MLWFCs when the surrounding octahedra are tilted, as illustrated by the snapshot in Fig.~\ref{fig:collective_rotor_cs}d.
The simulation snapshot shows a representative structural arrangement where Br-MLWFCs point toward the Cs when the lattice distortion results in decreased Br-Cs distances (pink dashed cylinders).
Similarly, Cs-MLWFCs point toward Br with larger Br-Cs distances (cyan dashed cylinders), resulting in an arrangement one might expect for ordering of quadrupolar sites. 
We also note that the observed orientational ordering of the Cs-MLWFC unit does not necessarily involve a displacement of Cs.
Out of phase octahedral tilting and Cs displacement are partner functions to different irreps, $T_{1u}$ and $T_{2u}$, respectively, and as a result they are not coupled. 
Instead, it is the distortions of the surrounding Sn-Br cube, which are a manifestation of out-of-phase octahedral tilting, that couple to the observed local electronic fluctuations of Cs.
These results suggest that the $A$-site cation can play a role in the electronic and lattice fluctuations of halide perovskites.

\section{Conclusion}
In summary, we used order parameters rooted in group theory to perform a statistical dynamic analysis of the coupling between local electronic symmetry of ionic sites and vibrational fluctuations in CsSnBr$_3$.
We found that there is significant static and dynamic coupling of local electronic structure to ionic displacements,
including coupling of electronic symmetry between sites. 
Of particular interest is the coupling of electronic symmetry and the resulting orientation of localized electron density to specific phonon modes, and we demonstrated that the octahedral tilting mode of CsSnBr$_3$ is strongly coupled to the orientation of bromide electron density on timescales of 5-10~ps. 
Using our analysis, the emergence of an octahedrally distorted structure at low temperatures naturally emerges, and we described how these distortions lead to long lived orientations of the electronic density~\cite{fabini2024noncollinear}. 
Fluctuations in local electronic structure govern fluctuations of the conduction and valence band energies, as well as the band gap~\cite{remsing2020new}, such that our analysis may aid in understanding optical processes in halide perovskites, including coupling of tilts to polarons~\cite{miyata2017lead,guo2019dynamic,schilcher2021significance} and excitons~\cite{yazdani2024coupling}.
Similarly, we anticipate that the detail provided by our analysis of local electronic fluctuations will aid in building further connections between ``local polar fluctuations'' and their impact on processes like carrier transport~\cite{schilcher2023correlated,guo2019dynamic,mayers2018lattice,schilcher2021significance}.
Furthermore, it was recently shown that rotation of localized electron pairs can couple to ion diffusion, resulting in electronic paddle-wheels in solid-state ion conductors~\cite{dhattarwal2024electronic}, such that the understanding of electronic fluctuations provided here may aid in quantifying mechanisms of ion and defect mobility in halide perovskites~\cite{eames2015ionic,limmerpnas,dey2024substitution}.

Of particular interest is the coupling between the orientation of localized electron pairs and nuclear displacements identified here.
We suggest that the orientational changes of localized electron pairs leads to nuclear translation---electronic rotation coupling, analogous to the translation--rotation coupling that is well-known for molecular plastic crystals~\cite{lynden1994translation}.
In molecular plastic crystals, translation--rotation coupling impacts many properties of the solid, including softening of related phonon modes.
Building on this analogy, we anticipate that the nuclear translation--electronic rotation coupling identified here results in phonon softening and anharmonic vibrations in halide perovskites. 
Indeed, based on our analysis of symmetry-dependent pair correlations, we suggested that vibrational (phonon) modes are softened as a result of their coupling to localized electron density fluctuations.
Moreover, the general ideas presented here are not limited to halide perovskites, and we anticipate that couplings between the orientation of localized electron density and nuclear displacements will be ubiquitous in anharmonic solids with localized electrons~\cite{laurita2022}.
%

\begin{acknowledgements}
We thank Harender Dhattarwal and Karin Rabe for helpful discussions. 
We acknowledge the Office of Advanced Research Computing (OARC) at Rutgers,
The State University of New Jersey
for providing access to the Amarel cluster
and associated research computing resources that have contributed to the results reported here.
\end{acknowledgements}

\bibliography{references}  

\end{document}